\documentclass[twocolumn,twoside,showkeys]{revtex4-1}

\usepackage{txfonts}
\usepackage[ansinew]{inputenc}
\usepackage[T1]{fontenc}
\usepackage{mathrsfs}
\usepackage{textcomp,booktabs}
\usepackage{graphicx}
\usepackage[hang]{subfigure}
\usepackage{multirow}
\usepackage{natbib}

\bibpunct{(}{)}{;}{a}{}{,} 

\usepackage{color}

\def\bval{B = \left(124\,{}\pm6\,\mathrm{(stat.)}\,{}^{+15}_{-6}\,\mathrm{(sys.)}\right)\,\mu\mathrm{G}}
\def\bfield{$\bval$}

\newcommand{\Diff}[2]{\frac{\mathrm{d} #1}{\mathrm{d} #2}}

\newcommand{\Difft}[2]{\mathrm{d} #1 \slash \mathrm{d} #2}

\newcommand{\unit}[1]{\,\mathrm{#1}}

\newcommand{\Nel}{N_\mathrm{el}}

\begin{document}
   \title{The Crab Nebula as a standard candle in \\
 very high-energy astrophysics}

   \author{M. Meyer,\footnote{\texttt{mmeyer@physik.uni-hamburg.de}}
          D. Horns, 
	  H.-S. Zechlin
	  }

   \affiliation{Institut f\"ur Experimentalphysik, University of Hamburg,
              Luruper Chaussee 149, D-22761 Hamburg, Germany\\
         }
   \date{\today -- Accepted in \textit{Astronomy \& Astrophysics}, July 18, 2010 }

  \begin{abstract}
	
The continuum high-energy gamma-ray emission between $1$~GeV and $10^5$~GeV
from the Crab Nebula has been measured for the first time in overlapping energy
bands by the Fermi large-area telescope (Fermi/LAT) below $\approx 100$~GeV and
by ground-based imaging air Cherenkov telescopes (IACTs) above $\approx
60$~GeV. To follow up on the phenomenological approach suggested by Hillas et
al. (1998), the broad band spectral and spatial measurement (from radio to
low-energy gamma-rays $<1$~GeV) is used to extract the shape of the electron
spectrum. While this model per construction provides an excellent description
of the data at energies $<1$~GeV, the predicted inverse Compton component matches
the combined Fermi/LAT and IACT measurements remarkably well after including
all relevant seed photon fields and fitting the average magnetic field to
\bfield. The close match of the resulting broad band inverse Compton component
with the combined Fermi/LAT and IACTs data is used to derive instrument
specific energy-calibration factors. These factors can be used to combine data
from Fermi/LAT and IACTs without suffering from systematic uncertainties
on the common energy scale. 
 As a first application of the cross calibration, we
derive an upper limit to the diffuse gamma-ray emission between 250~GeV and 1~TeV
based upon the combined measurements of Fermi/LAT and the H.E.S.S. ground-based
Cherenkov telescopes. Finally, the predictions of the magneto-hydrodynamic flow model of
Kennel \& Coroniti (1984) are compared to the measured SED.
\end{abstract}


   \maketitle
%

\section{Introduction}
The Crab Nebula 
has been and remains
an intensely studied object in
astrophysics \citep[for a recent review see e.g.][]{2008ARA&A..46..127H}. It is part of the remnant of a core-collapse supernova that occurred in
1054 AD at a distance of $d \approx 2 \unit{kpc}$ \citep{1968AJ.....73..535T}.
Observations of the nebula have been carried out at every accessible wavelength
resulting in a remarkably well-determined spectral energy distribution (SED).
Therefore, the Crab Nebula is an ideal object for detailed studies of 
the conversion of Poynting flux to particle energy flux \citep[see e.g.][]{1990ApJ...349..538C,2003ApJ...591..366K,2008AIPC..983..200A}
and finally of the
acceleration processes taking place at the termination shock \citep[see e.g.][]{1987ApJ...321..334E,2004ApJ...603..669S}.
The commonly considered model for the Crab Nebula \citep[see e.g.][henceforth
RG74]{1974MNRAS.167....1R} assumes an ultra-relativistic outflow from the
pulsar that terminates in a standing shock at a distance $r_s$ which is roughly  10\%
of the total nebula's size. In the downstream medium, the particles are pitch-angle isotropized, forming a broad power law in energy. The
magneto-hydrodynamic (MHD) analysis of the downstream flow by
\citet{1984ApJ...283..694K}, henceforth KC84, provides an elegant solution to
the particle distribution and the magnetic field downstream of the shock under
the assumption of a particular \textit{ad hoc} injection spectrum.  In the
framework of this model, the radio emission is explained by a separate electron
population, which has been linked to the high spin-down phase of the pulsar
\citep[relic electrons, ][]{1999A&A...346L..49A} or by acceleration in MHD
turbulences \citep{2004A&A...423...13N}. 

Several different models have been proposed to explain the observed high-energy gamma-rays as inverse Compton emission from the same electron
population responsible for the synchrotron X-ray emission (see, e.g., \citealt{1992ApJ...396..161D,1996MNRAS.278..525A,1998ApJ...503..744H,2003A&A...405..689B,2004ApJ...614..897A,2008ApJ...676.1210Z, 2008A&A...485..337V}. 
  We point out that \citeauthor{2008A&A...485..337V} use a time-dependent axisymmetric numerical simulation of the nebula's evolution).
The large uncertainties in the observational data in the past have only weakly
constrained the models at high energies, i.e., between 1~GeV and 100~GeV.
The largely improved statistics of the measurements carried out with
the recently commissioned Fermi/LAT \citep{2009ApJ...697.1071A} have provided us
with more accurate data in this crucial energy window. Nevertheless,
the combined observations carried out with different instruments in overlapping
energy bands (Fermi/LAT and IACTs) are currently limited by the systematic
uncertainties of the relative and absolute energy calibration. 

Because of this, a two-pronged approach is followed here. On the
one hand, we perform accurate modeling of the available measurements of the broad band SED of the Crab Nebula. On the other, the model is used to
derive corrections of the measured energy scale for the individual instruments
to a common energy scale. This cross calibration reduces the systematic
uncertainties and proves useful for any study which relies on the combination 
of spectral measurements of the Fermi/LAT and IACTs.

 Given that we are mainly interested in the spectral modeling of the high-energy emission, which is currently not spatially resolvable, we primarily
 apply a simple and robust approach based upon the work of
\citet{1998ApJ...503..744H}, which we refer to as the constant B-field model.  Nevertheless, we also investigate the MHD flow model 
suggested by KC84 and its application to the 
inverse Compton component \citep[][henceforth AA96]{1996MNRAS.278..525A}.

The article is organized as follows. In Section \ref{sec:SED} the available data and modeling of the SED
are presented and discussed. Both, the simplified constant B-field model and the MHD flow model of KC84 
are fitted to the measured synchrotron part of the SED by means of a $\chi^2$-minimization.
The results of the two models are compared.
 In Section \ref{sec:el_spec}, we discuss the properties
of the underlying  electron spectrum of the simplified model as derived from the observations. The
results of a first instrumental cross calibration are presented in Section
\ref{sec:crosscal}, together with an application to extract limits on the
diffuse $\gamma$-ray background at TeV energies. 

\section{Spectral energy distribution of the Crab Nebula}
\label{sec:SED}
The compilation of observational data used here is summarized by
\citet{2004ApJ...614..897A} and references therein. The radio data \citep{2010ApJ...711..417M}
have been corrected for a secular decline of $-0.18~\%/$yr to a common date
(01/01/2000).  Additionally, new data are added and listed in Table \ref{tbl:data},
while the entire compilation of data is displayed in Fig. \ref{fig:SED:constant} and \subref{fig:SED:mhd}. The far
infrared (FIR) observations from Spitzer, ISO, and Scuba deviate from a simple
power-law extrapolation of the radio spectrum.  The flux of the optical line
emission is taken from \citet{1985ARA&A..23..119D}, \citet{1987AJ.....94..964D}
and \citet{1990ApJ...357..539H}. The optical line emission of the filaments in
the nebula is estimated in the following way. The high-resolution spectral
observations of individual filaments have been corrected for extinction
\citep{1990ApJ...357..539H} and scaled to match the global emission from the
filaments \citep[see the discussion in][]{1985ARA&A..23..119D}.

New X-ray observations by the XMM-Newton and INTEGRAL observatories have been
chosen to replace older measurements. Both, the XMM-Newton and INTEGRAL (with
the instruments SPI and IBIS/ISGRI) observatories are calibrated on the basis
of detailed simulations and laboratory measurements \citep{1998SPIE.3444..278G,1998SPIE.3444..290G,2003A&A...411L..71A}.
 This approach differs from
commonly used methods in which corrections of the instrument's response
function are applied to reproduce a specific spectral shape and flux
of the Crab Nebula. As a result, corrected measurements are model-dependent
and therefore we chose not to include them here. The cameras of XMM-Newton
spatially resolve the Crab Nebula, whereas the measurements of SPI and
IBIS/ISGRI may include contributions from the pulsar, possibly leading to
higher fluxes in comparison to the XMM-Newton observations. 
We stress that the
difference in flux normalization between XMM-Newton and SPI are beyond the
systematic errors quoted. To combine the two measurements, we chose
to scale the flux of the SPI data by a factor of $0.78$ to match the extrapolation of the measurement of XMM-Newton. 

\begin{table}
\centering
\caption[Observations for SED of the Crab]{Observations used for the SED.
All other data shown in Fig. \ref{fig:SED:constant},\subref{fig:SED:mhd} are taken from \citet{2004ApJ...614..897A} and references therein.
(1) {\citet{2004MNRAS.355.1315G}};
(2) {\citet{2006AJ....132.1610T}};
(3) {\citet{2005SPIE.5898...22K}};
(4) {\citet{2009ApJ...704...17J}};
(5) {\citet{2008int..workE.144J}};
(6) {\citet{2010ApJ...708.1254A}};
(7) {\citet{2006A&A...457..899A}};
(8) {\citet{2008ApJ...674.1037A}}
}
\begin{tabular}{llc}
\textbf{Energy Band} & \textbf{Instrument} & \textbf{Reference}\\
\hline
\hline
Submillimeter		& ISO \& SCUBA			& (1)  \\
 to far infrared	& SPITZER			& (2) \\
 \hline 
 X-ray to 		& XMM-Newton			& (3)\\
 $\gamma$-ray		& SPI				& (4)\\
 {}			& IBIS$\slash$ISGRI 		& (5)\\
 {}			& Fermi / LAT			& (6)\\
 \hline
 VHE			& H.E.S.S.			& (7)\\
 {}			& MAGIC				& (8)
\end{tabular}
\label{tbl:data}
\end{table}

\subsection{Constant B-field model}
\label{subsect:constB}
The nebula is assumed to be filled with relativistic electrons with an averaged
total differential number $\Difft{N_\mathrm{el}}{\gamma}$.
 To explain the change in the continuum between the radio and infrared
(see above),
we distinguish between radio and wind electrons.
 The two
spectra are assumed to follow power laws with appropriate cut-offs.
For the radio-emitting electrons, a power law 
between $\gamma^r_\mathrm{min}=22$ (a value only constrained towards higher values of $\gamma$ and kept constant in the fit),
 and $\gamma^r_\mathrm{max}$
\begin{eqnarray}
\Diff{\Nel^r}{\gamma} &=& \left\{\begin{array}{ll} N_0^r \gamma^{-S_r} & \mathrm{for}\quad \gamma^r_\mathrm{min}\le\gamma\le\gamma^r_\mathrm{max},\\ \\
0 & \mathrm{otherwise},\end{array}\right.\label{eqn:radio_n}
\end{eqnarray}
is sufficient, while for the 
wind electrons a broken power law with a superexponential cut-off  at the lower 
end is required:
\begin{eqnarray} 
\Diff{\Nel^w}{\gamma} &=&  N_0^w\left\{\begin{array}{ll}
\left(\frac{\gamma}{\gamma_\mathrm{break}^w}\right)^{-S_w},
&\mathrm{for}\quad\gamma<\gamma_\mathrm{break}^w, \nonumber\\
\\
\left(\frac{\gamma}{\gamma_\mathrm{break}^w}\right)^{-(S_w+\Delta S)},
& \mathrm{for}\quad\gamma^w_\mathrm{break}\le\gamma\le\gamma_\mathrm{max}^w ,\\
\\
0, & \mathrm{for}\quad \gamma > \gamma^w_\mathrm{max}, \\
\end{array}\right\}\nonumber\\
&{}&\quad\times\exp\left(-\left[\frac{\gamma_\mathrm{min}^w}{\gamma}\right]^{\beta}\right)\label{eqn:wind_n}.
\end{eqnarray}

As assumed by \citet{1998ApJ...503..744H}, the electron-number volume density $n_\mathrm{el}^w(\gamma,r)$ 
in the nebula is taken to drop off radially following
a Gaussian function such that   
\begin{eqnarray}
  n_\mathrm{el}^w(\gamma,r) &=& \frac{n^w_0}{N_0^w} \exp\left[-\frac{r^2}{2\rho^2(\gamma)}\right] \Diff{\Nel^w}{\gamma}.
\end{eqnarray}
The width of the Gaussian $\rho(\gamma)$ decreases with
increasing Lorentz factor $\gamma$ to account for the observed shrinking of the
nebula (see Appendix~\ref{sec:seed} for further details). The volume of the
nebula is assumed to be filled with an entangled magnetic field of constant
field strength. Within the nebula, various seed photon fields are upscattered
by the electrons via the inverse Compton process.  The effective density of seed
photons $n_\mathrm{seed}$ is calculated by convolving the electron density with
the photon density and summing the contributions from several different
components: (1) synchrotron radiation, (2) emission from thermal dust, (3) the
cosmic microwave background radiation (CMB), and (4) optical line emission from
the nebula's filaments. Similar to the electron population, the spatial photon
densities are approximated by (energy-dependent) Gaussian distributions, whereas
the photon density of the CMB is assumed to be constant throughout the nebula. The
relevant relations for the seed photon fields are summarized in Appendix
\ref{sec:seed} \citep[for further details see also][]{1998ApJ...503..744H}. The resulting synchrotron and inverse Compton emission (at
frequency $\nu$) is found by \begin{eqnarray}
\label{eqn:emission}
 L_\nu &=& \int\limits_1^\infty\mathrm{d}\gamma  \Diff{\Nel}{\gamma} \left (\mathcal{L}_\nu^\mathrm{Sy} + 
\mathcal{L}_\nu^\mathrm{IC}\right).
\end{eqnarray}
The specific single particle emission functions $\mathcal{L}_\nu$ for
synchrotron (Sy) and inverse Compton (IC) processes are given in Appendix~\ref{Appi}.

For a given value of the magnetic field, the shapes of the two electron spectra
are varied until the resulting synchrotron spectrum
reproduces the observational data. 
 The best-fit values for the ten free parameters describing the electron spectrum at a fixed magnetic field strength 
are determined by means of a least-squares algorithm
(Levenberg-Marquardt), which provides the closest match of the expected
synchrotron spectrum (between $1$ and $10^{14}$~GHz) with the observational data (see Table~\ref{tbl:el_spec} for the
results of the fit and the discussion in Section \ref{sec:el_spec}).
The resulting $\chi^2/\mathrm{d.o.f}=214.5/217$ 
is below unity after a relative systematic uncertainty of $7$~\% 
was added in quadrature to the statistical error quoted for the data in the 
literature. The minimization procedure is not sensitive to the particular choice of starting values and it converges
reliably.
The resulting covariance matrix allows for an analysis of the correlations
between various parameters.
As expected, the normalizations are anti-correlated to
the power law indices. Additionally, there are modest anti-correlations between
the position of the break in the spectrum of the wind-electrons with the power-law indices.
The matrix of correlation coefficients is listed in Appendix~\ref{App:Cov}.

The predicted inverse Compton component above 700~MeV is then used to calculate $\chi^2(B)$ for a range of $B$-field values
using only the Fermi data. The best-fit value of $B$ and its
statistical uncertainties are estimated by adjusting a parabola to $\chi^2(B)$
and calculating its second derivative. After taking
the systematic energy uncertainty on the global energy scale of the Fermi/LAT
data into account, $\Delta
E/E = {}^{+5\%}_{-10\%}$ \citep[see e.g.][]{2009PhRvL.102r1101A}, the average
$B$-field is found to be 
\begin{equation} 
\bval\label{eqn:Bfit}.
\end{equation}
The reduced $\chi^2/\mathrm{d.o.f.}=6.37/13\approx 0.49$ for the fit of the inverse
Compton component to the Fermi data indicates that the statistical 
errors on the Fermi differential flux may be slightly overestimated.  
\\

\begin{figure*}[t!]
 \centering
 \subfigure[
The SED of the Crab nebula calculated
in the framework of the constant B-field model. The open blue data points have been included in the fit
for the synchrotron part and the filled blue points used to determine the best-fitting magnetic field. ]
{
 \label{fig:SED:constant}
 \includegraphics[angle=270,width = 0.90\textwidth]{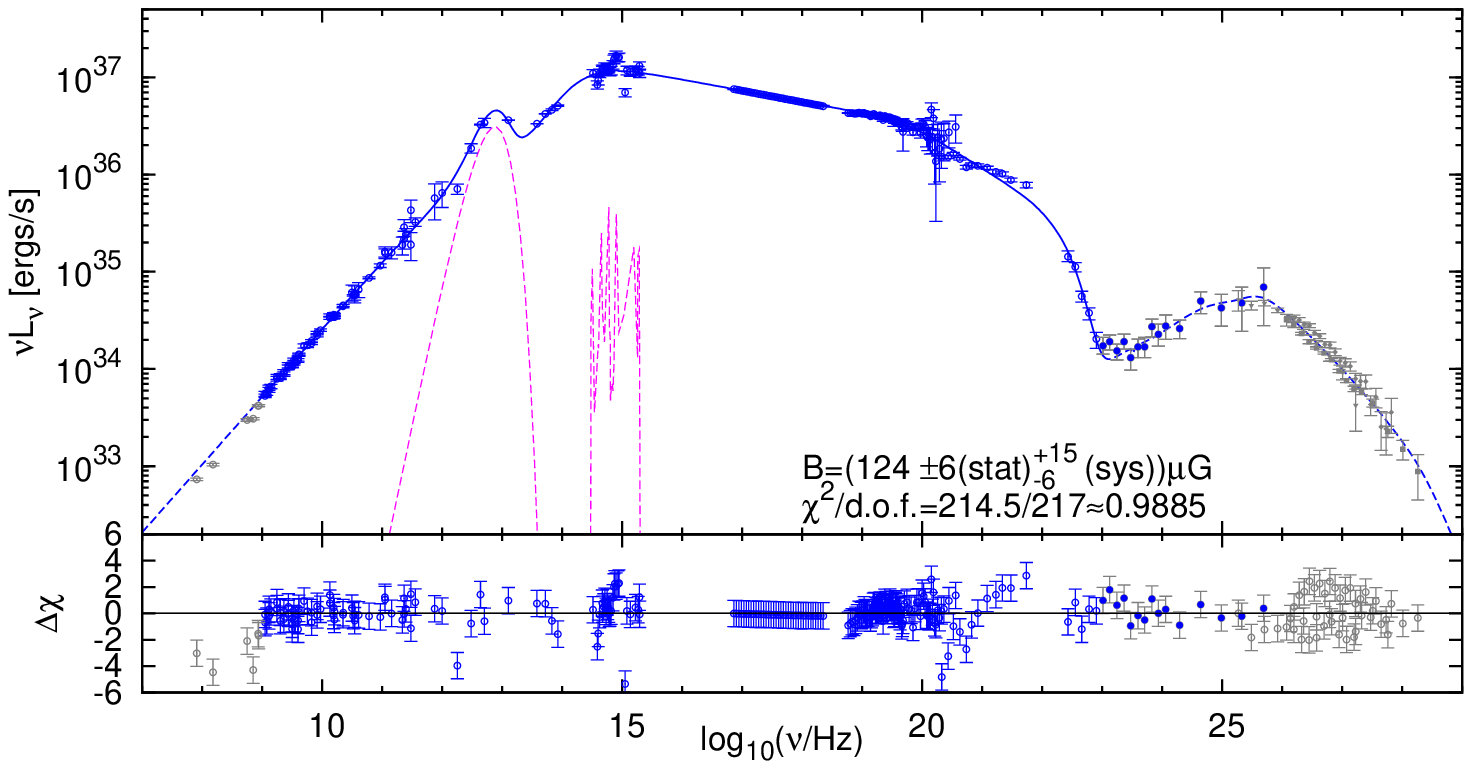}
}
 \centering
\subfigure[
The SED of the Crab nebula (data identical to Fig.~\ref{fig:SED:constant}) with a best-fit model using
an MHD flow description of the pulsar wind nebula.]
{
 \label{fig:SED:mhd}
 \includegraphics[angle=270,width = 0.90\textwidth]{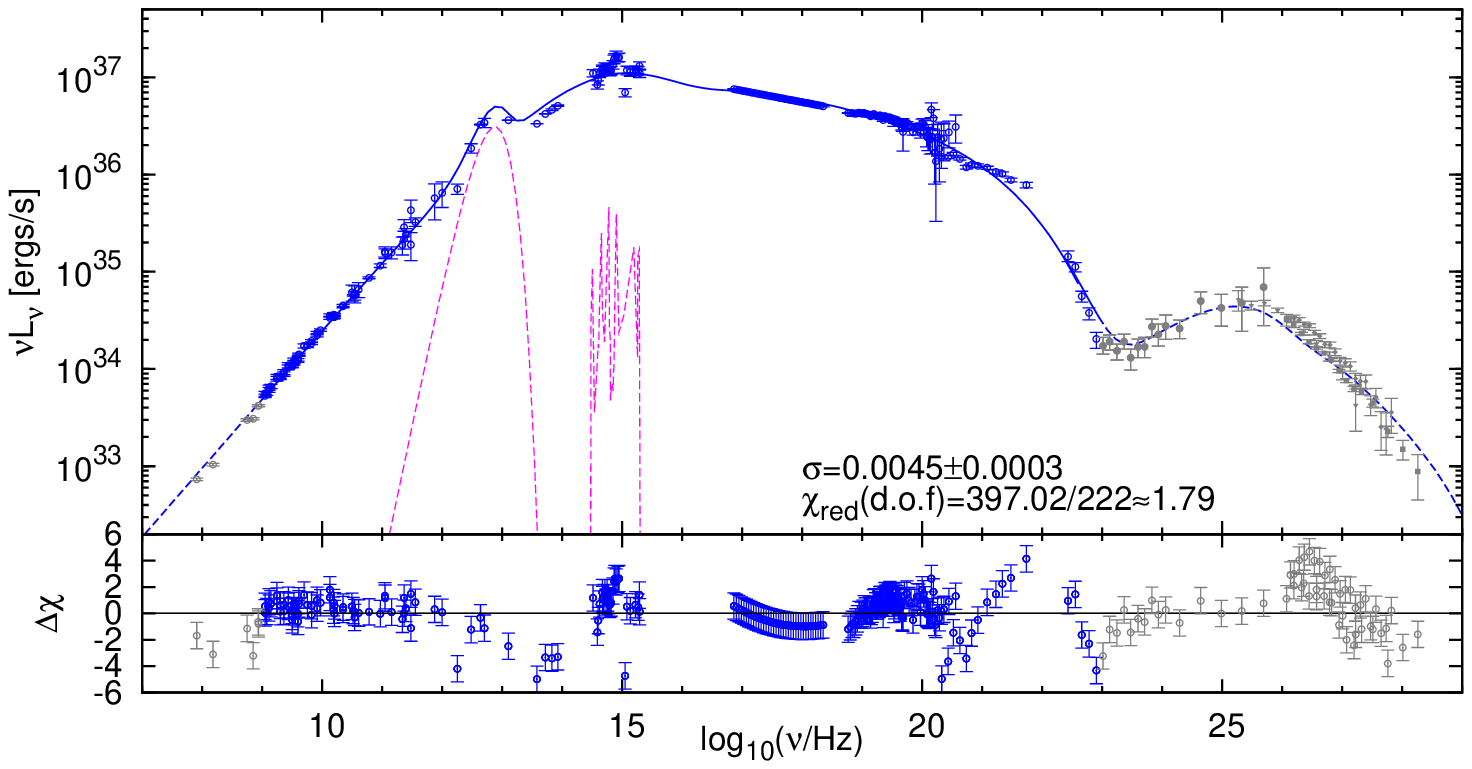}
}
\caption[]{The SEDs for the constant B-field and MHD flow model. The data
compilation is based upon \citet{2004ApJ...614..897A} with updates listed in
Table~\ref{tbl:data} and modifications (e.g. scaling of the SPI flux) described in
the text.  See Section \ref{sec:SED} for further details.
The synchrotron emission fitted to the data is shown by the blue solid line while the remaining components (low energy part and inverse Compton emission) naturally result from the model.
Therefore, blue data points have been included in the fit whereas the gray points have been excluded. 
In addition to the nonthermal
continuum emission, the contribution of dust and line emission from the filaments is indicated with
a magenta dashed line. The lower panels show the residuals of the fit.
.}
\end{figure*}

\begin{figure*}[t!bh!]
\centering
\includegraphics[width=0.9\textwidth]{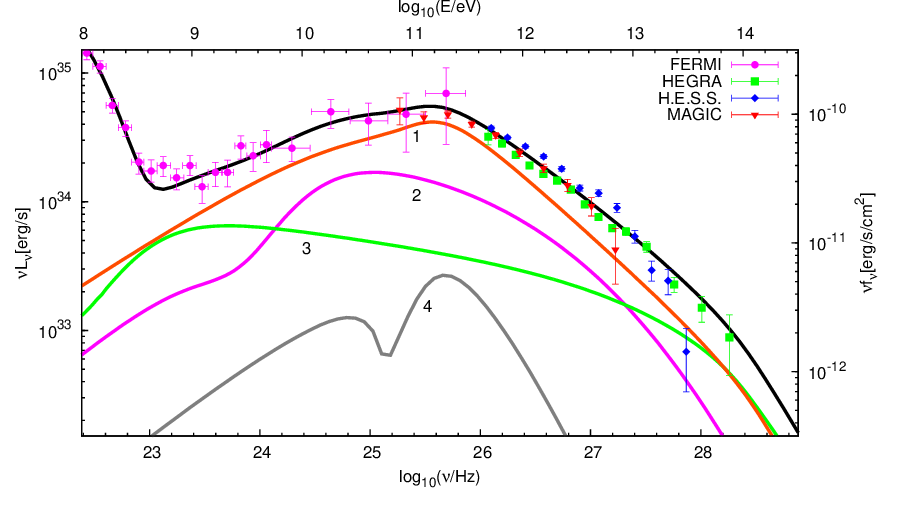}
 \caption{\label{fig:IC} The Fermi/LAT and IACT data points are shown along with the 
total IC flux of the constant B-field (black line) and individual components from 
the different seed photon fields (constant B-field model):  (1) synchrotron, (2) thermal dust, (3) CMB, and (4) line emission from filaments.  }
\end{figure*}

The resulting broad band SED is displayed in Fig. \ref{fig:SED:constant} (
solid and dashed blue curves) including synchrotron and IC emission and thermal
emission from the dust in the nebula and optical line emission from the filaments
(dashed magenta line). The inverse Compton component including the various
contributions of different seed photon fields is shown in detail in
Fig.~\ref{fig:IC}. For convenience, an analytical parametrization of the
derived energy flux above $1$~GeV is presented in Appendix \ref{sec:app-para}.
This parametrization is especially useful to derive cross calibration factors
as introduced in Section \ref{sec:crosscal} for other or future instruments
that measure the same energy range. 

For the thermal dust emission a graybody spectrum is used.  By fitting  the
combined spectrum (thermal and nonthermal emission) to the data a temperature
of $T = 93 \unit{K}$ is derived.  The graybody peaks at $(\nu
f_\nu)_\mathrm{max} \approx 6.45\times10^{~-9}
\unit{ergs}\unit{s}^{-1}\unit{cm}^{-2}$.  With the relation
\citep{1998PASP..110....3G}
\begin{equation}
M_\mathrm{dust} = \frac{1.36(\nu f_\nu)_\mathrm{max}~a \rho ~4\pi d^2}{3\sigma_\mathrm{SB} T^4},
\end{equation}
the dust mass $M_\mathrm{dust}$ can be estimated. Here $\sigma_\mathrm{SB}$ denotes the Stefan-Boltzmann constant. 
We adopt the same assumptions as \citet{2006AJ....132.1610T}; i.e., we assume a dust grain size of $a = 10\,\mu\mathrm{m}$ and 
graphite grains with a density of $\rho = 2.25 \unit{g}\unit{cm}^{~-3}$. This leads to a dust mass of $M_\mathrm{dust} = 4\times10^{-4} M_\odot$,
about 40\% of the value obtained by \citet{2006AJ....132.1610T}.

\subsection{MHD flow model}
 While the treatment presented above provides an accurate match of the measured broad band SED, 
a more physical description of the evolution of the injected particles and magnetic field 
in a spherical volume has been suggested by KC84. 
In addition to the synchrotron component already studied
by KC84, AA96 extended this approach by investigating the high-energy (inverse Compton) component of the SED. The MHD solution requires the
shock distance $r_s$, the magnetization parameter $\sigma$, i.e. the ratio of
the Poynting flux to the particle energy flux at the position of the
shock, and the shape of the injected electron spectrum (particle number per unit volume $n(\gamma)$ 
in the interval $\gamma$ to $\gamma+d\gamma$) as main input parameters. 

Previously, RG74 assumed a shock distance of $r_\mathrm{s}=0.10~\unit{pc}$, 
which in the meantime has become the canonical value, whereas recent high spatial resolution Chandra observations 
\citep{2000ApJ...536L..81W} indicate that the
shock (if identified with the bright inner ring in the X-ray image at $14"$ distance to the pulsar) 
 resides at a distance of $r_\mathrm{s} = (0.14\pm 0.01) \unit{pc}$. 

Based upon the initial analysis of KC84, the magnetization paramater was considered
to be less than 1\% with typical values ranging from $\sigma=0.001$ to
$\sigma=0.005$. With such low values of $\sigma$,
  both the observed break in
the spectrum between optical and radio wavelengths and the general
morphology of the nebula are reproduced for the most part. 
However, the asymmetry in the brightness of the far and near sides of the torus
observed in X-rays and the measured flow speed \citep{2004ApJ...609..186M}
imply a higher value of $\sigma$ between $0.01$ and $0.13$, more likely 
close to $\sigma=0.05$ \citep{2003MNRAS.346..841S, 2004ApJ...609..186M}.

Following the approach suggested by AA96, the broad band SED is calculated (see Eq.~\ref{eqn:emission}) 
including
the same  external seed photon fields as described above and two populations of electrons. 
The  radio-emitting relic electrons are  distributed homogeneously in the nebula with
\begin{eqnarray}
n_\mathrm{radio}(\gamma)=n_0\gamma^{-S_r}~\exp\left(-\frac{\gamma}{\gamma_\mathrm{max}^r}\right),
\label{eqn:mhd_radio}
\end{eqnarray}
 whereas a population of wind electrons is injected at the shock:
\begin{eqnarray}
n_\mathrm{wind}(\gamma)&=&
 q_0 (\gamma+\gamma^w_\mathrm{min})^{-S_w}\exp\left(-\frac{\gamma}{\gamma_\mathrm{max}^w}\right). 
 \label{eqn:mhd_wind}
\end{eqnarray}
The radiative and adiabatic cooling of the wind electrons is treated in the same way
as suggested by AA96.
The best-fitting value for the seven parameters describing the electron spectra 
are found in a similar way to what is described above. 
For a fixed value of $\sigma$, the parameters
describing the electron distribution ($n_0$, $S_r$, $\gamma_\mathrm{max}^r$, $q_0$, $\gamma^w_\mathrm{min}$, $S_w$, $\gamma_\mathrm{max}^w$)
are  varied, until the predicted synchrotron emission is matched best to the same data as used for the constant B-field model. 
The procedure is repeated for a range of values of $\sigma$ until an absolute minimum is found at $\sigma=0.0045\pm0.0003$
and $\chi^2/\mathrm{d.o.f.}=397.02/222\approx 1.79$ with a probability of obtaining a higher value of $\chi^2$ per chance $P(>\chi^2)=4.8\times 10^{-12}$
(see Table~\ref{tbl:el_spec_mhd} for the best-fit values and Appendix~\ref{App:Cov} for the correlation coefficients). 
The high $\chi^2$ value indicates that the simple MHD-flow
model fails to describe the synchrotron part of the SED in detail, while the overall shape
is certainly correct. It is noteworthy that
the high value of $\gamma^r_\mathrm{max}$ implies that most of the optical emission is produced by the same population of electrons as are
responsible for the radio emission. In this case, the spatial extension predicted at optical frequencies would be similar to the 
extension of the radio nebula which clearly contradicts observations. When looking at the inverse Compton component calculated
in this model, the agreement between the Fermi part of the spectrum and the model is fairly good, even though the shape of the synchrotron
cut-off is harder than the measured spectrum. At energies beyond the position of the peak in the inverse Compton part of the SED, the
model spectra are considerably harder than the actual measurements. The discrepancy is directly related to the mismatch that is evident
in the X-ray part of the synchrotron spectrum. 

\begin{figure}[bh!]
 \centering 
 \includegraphics[angle = 270, width = 0.48\textwidth]{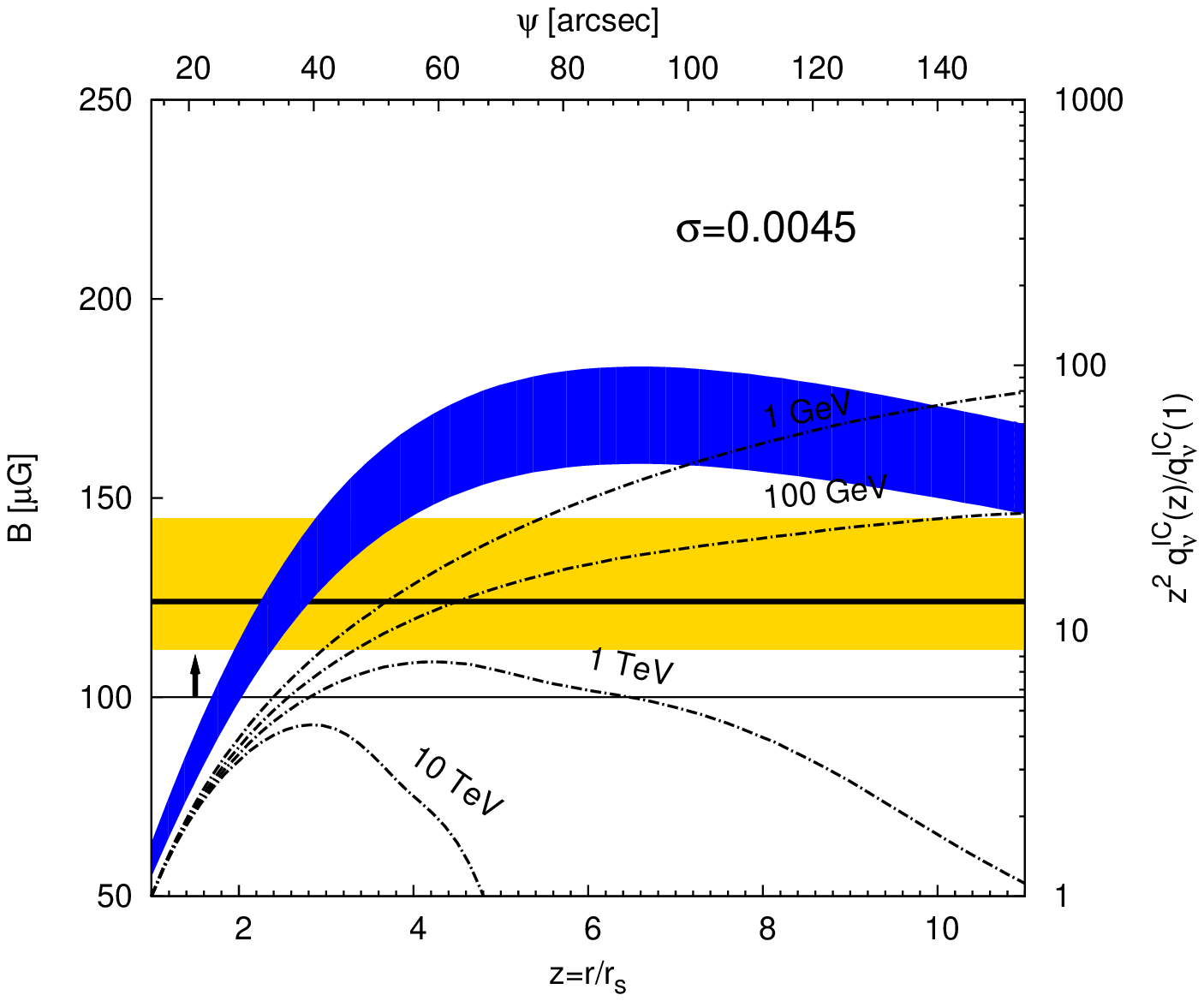}
 \caption[]{
The black horizontal line and the yellow shaded region mark the volume averaged
magnetic field and its uncertainties as derived from the constant B-field model (see Eq. \ref{eqn:Bfit}). 
The magnetic field configuration as derived from the MHD solution of the downstream flow (KC84)
is shown for  $\sigma=0.0045$  for a shock distance between
$0.13~\mathrm{pc}<r_s<0.15$~pc, where $r_s=0.13$~pc results in the upper and $r_s=0.15$~pc in the lower boundary, respectively. The
dashed-dotted lines mark the relative emissivity (normalized to $z=1$) at specific energies. The horizontal line at $100~\mu$G indicates a 
lower limit on the $B$-field, see Section \ref{sec:el_spec}.
}
\label{fig:BMHD}
\end{figure}

\subsection{Comments on the two different model approaches}

The average magnetic field derived above (see Eq.~\ref{eqn:Bfit}) is displayed
as a yellow band in Fig.~\ref{fig:BMHD} together with the MHD solution of the
magnetic field for  $\sigma=0.0045$.
It is obvious that, in a spatially varying magnetic field, the effective magnetic field seen by the population of electrons
radiating at different gamma-ray energies varies as well. Therefore, it should  be possible to determine $\sigma$ from the combined spectral
measurement of the synchrotron and inverse Compton components. However, it
appears that the broad band spectrum in the KC84 model is not consistent with
the data, which implies that some of the assumptions may have to be refined
before using the model further in a quantative way in combination with the improved
data available.

A straightforward extension of the spherical KC84 model would be to incorporate an
asymmetric flow with a modulated  magnetization \citep[see e.g.][]{2004MNRAS.349..779K,2004A&A...421.1063D,2006A&A...453..621D,2008A&A...485..337V}.
Depending on the particular way the magnetization just upstream of the shock varies, the
superposition of the emission from regions with a different downstream flow
magnetization could be arranged to  be closer to the observations than the
single $\sigma$-model of KC84. In fact, by e.g. superposing the emission of two separate regions with $\sigma\approx 0.01$ and 
$\sigma\approx 0.001$ would provide a better overall description of the data.
Given that the resulting volume-averaged
magnetic field in the downstream region could approach an effectively constant field,
the resulting inverse Compton emission would be comparable to the
simple model considered here.

\section{Electrons in the nebula}
\label{sec:el_spec}
The total number electron spectrum $\Difft{\Nel}{\gamma}$ of the constant B-field model is shown in Fig.~\ref{fig:el_spec}. 

\subsection{Radio electrons}
\label{sec:radioel}
\begin{figure*}[t!bh]
 \centering
 \includegraphics[width = 0.9\textwidth]{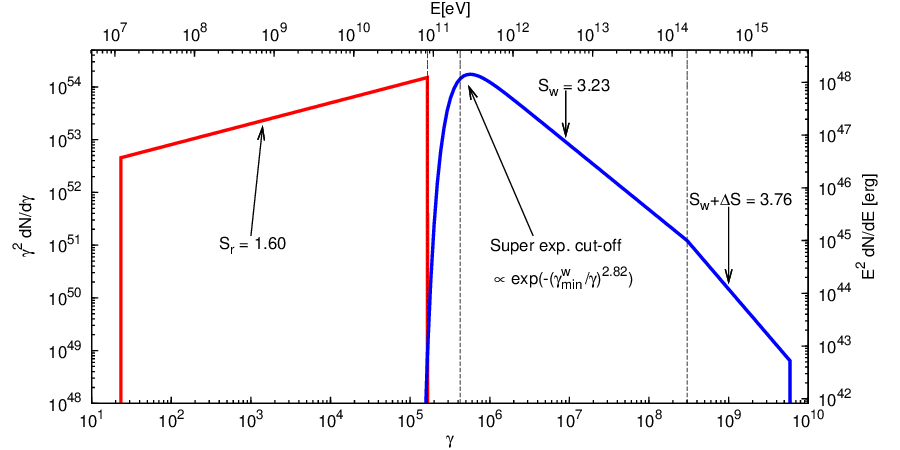}
 \caption{The two components of the electron spectrum used to calculate the broad band emission of the Crab Nebula in the constant B-field model.
Red solid line: radio electrons; blue solid line: wind electrons. The black dashed lines indicate the values fot the minimum, maximum, and break energies.}
 \label{fig:el_spec}
\end{figure*}

\citet{1999A&A...346L..49A} has suggested that the radio-emitting electrons (see Section \ref{subsect:constB} and Fig. \ref{fig:el_spec}) were injected in the phase of rapid
spin-down during the initial stages of the pulsar-wind evolution.
 Observations of time-variable emissions from radio wisps with a hard spectrum indicate ongoing acceleration of
radio emitting electrons \citep{1990ApJ...357L..13B,1993ApJ...416..251K,2004ApJ...615..794B}.
However, the
injection rate determined from the observations of the wisps is not high enough to explain the 
total population of radio-emitting electrons, therefore we consider them to be relic electrons.

\begin{table*}[t!hb]
\caption{Parameters for the electron spectrum used for the constant B-field model.
    At the energy $\gamma_\mathrm{min}$ the radio electron spectrum cuts off sharply whereas the wind electrons cut off superexponentially.
The energy $\gamma_\mathrm{max}$ denotes a sharp cut-off for the radio and wind electrons.}
\label{tbl:el_spec}
\centering
 \begin{tabular}{lc||c|c}
 \centering
\multirow{3}{*}{\textbf{Parameters}} &{}	& \multicolumn{2}{|c}{\textbf{Constant $B$-field model}} \\
{}	&{}&Radio		&Wind		\\	
\hline
\hline
Normalization constant\ldots\ldots\ldots\ldots\ldots\ldots\ldots\ldots.~.&$\ln(N_0)$	&$120.0(1)$ & $78.6(3)$ \\
Low energy cut-off\ldots\ldots\ldots\ldots\ldots\ldots\ldots\ldots\ldots\ldots.&$\ln(\gamma_\mathrm{min})$&$3.1$&$12.96(3)$\\
Super exponential cut-off parameter\ldots\ldots\ldots\ldots\ldots& $\beta$ & -- & $2.8(4)$\\
Break position \ldots\ldots\ldots\ldots\ldots\ldots\ldots\ldots\ldots\ldots\ldots.&$\ln(\gamma_\mathrm{break})$&--&$19.5(1)$\\
High energy cut-off\ldots\ldots\ldots\ldots\ldots\ldots\ldots\ldots\ldots\ldots&$\ln(\gamma_\mathrm{max})$&$12.1(7)$& $22.51(3)$\\
Spectral index \ldots\ldots\ldots\ldots\ldots\ldots\ldots\ldots\ldots.~.&$S$	&$1.60(1)$		&$3.23(1)$ 	\\
Spectral index (after break)\ldots\ldots\ldots\ldots\ldots\ldots\ldots.~.& $S + \Delta S$ & -- & $3.76(3)$\\
 \end{tabular}
\end{table*}

\begin{table*}[t!hb]
\caption{Parameters for the electron spectrum for the MHD flow model.
    Compare Eqs. \ref{eqn:mhd_radio} and \ref{eqn:mhd_wind}.
}
\label{tbl:el_spec_mhd}
\centering
 \begin{tabular}{lc||c|c}
 \centering
\multirow{3}{*}{\textbf{Parameters}} &{}	&  \multicolumn{2}{|c}{\textbf{MHD flow model}} \\
{}	&{}&Radio		&Wind\\
\hline
\hline
Normalization constant\ldots\ldots\ldots\ldots\ldots\ldots\ldots\ldots.~.&$\ln (n_0), \ln (q_0)$	&$-11.41(2) $ &  $-1.32(1)$\\
Low energy cut-off\ldots\ldots\ldots\ldots\ldots\ldots\ldots\ldots\ldots\ldots.&$\ln(\gamma_\mathrm{min})$& --        &  $13.94(2) $\\
High energy cut-off\ldots\ldots\ldots\ldots\ldots\ldots\ldots\ldots\ldots\ldots&$\ln(\gamma_\mathrm{max})$&$ 13.55(3)$&  $22.60(1)$ \\
Spectral index \ldots\ldots\ldots\ldots\ldots\ldots\ldots.~.&$S$	       &$1.58(1)$ 	         &$  2.32(1)$		\\
Magnetization parameter\ldots\ldots\ldots\ldots\ldots\ldots & $\sigma$ & \multicolumn{2}{c}{$0.0045(3)$} \\
 \end{tabular}
\end{table*}

The radio electrons loose energy radiatively and adiabatically as the nebula 
expands. Interpreting the maximum Lorentz factor of the radio electrons  as induced only by radiative cooling (ignoring adiabatic losses),
an upper limit on the magnetic field can be derived:
\begin{equation}
 B \le 385 \left(\frac{\gamma}{1.74\times 10^5}\right)^{-1\slash2} \left(\frac{t_\mathrm{sy}}{956\unit{yr}}\right)^{-1\slash2}\,\mu\mathrm{G}. 
\end{equation}
A lower limit on the magnetic field can be estimated by varying the magnetic
field and the normalization $N_0^r$ of the radio electrons until the inverse Compton
component overshoots the Fermi data. The constraint on the magnetic field 
is estimated to be $B\ge 100~\mu$G. {The two limits are consistent with the value derived for the constant B-field model as well as with the 
prediction by the KC84 model.}

\subsection{Wind electrons}
The continuously injected wind electrons produce the bulk of the observed SED above sub-mm/FIR wavelengths via synchrotron emission. The spectrum is given in Eq.~\ref{eqn:wind_n}. The radiatively cooled spectrum of
the wind electrons  has a spectral index of $S_w = 3.23 = 2.23 + 1$, which can
be explained naturally by 1${}^\mathrm{st}$-order Fermi acceleration at an
ultrarelativistic shock with subsequent synchrotron cooling \citep[see
e.g.][]{1999JPhG...25R.163K}. The low-energy cut-off of the wind electrons is
found to be $\gamma_\mathrm{min}^w=4.24\times10^5$, which is anticipated in the model
suggested by KC84. It requires the average Lorentz factor $\langle
\gamma_w\rangle$ of the isotropized (downstream) electrons to be similar to
the bulk Lorentz factor $\gamma_*$ of the upstream electrons, $\langle
\gamma_w\rangle = 7.33 \times 10^5 \approx \gamma_*$
\citep{1984ApJ...283..694K,1980A&A....83....1K,1996A&AS..120C..49A}. 
An additional feature present in the hard X-ray spectrum, which shows a softening 
at $\sim 130\unit{keV}$, corresponds to a break with $\Delta
S=0.43$ at $\gamma_\mathrm{break}^w = 3.01\times 10^{8}$ in the electron spectrum. The origin of this
feature in the electron spectrum is very likely related to the injection/acceleration, 
 given that it can hardly be related to energy-dependent escape. (The X-ray emitting electrons
suffer cooling well before escaping the nebula.) The value of $\Delta S=0.43$ could
hint at an energy dependent effect similar to diffusion in a Kolmogorov-type turbulence power spectrum. 

The total energy of the radio and wind electrons, respectively, is found to be
\begin{eqnarray}
 E_r & = & mc^2 \int\limits_{1}^{\infty} \,\gamma \Diff{\Nel^r}{\gamma}\,\mathrm{d}\gamma = 3.10\times10^{48}\unit{ergs},\\
 E_w & = & mc^2 \int\limits_{1}^{\infty} \,\gamma \Diff{\Nel^w}{\gamma}\,\mathrm{d}\gamma =2.28\times10^{48}\unit{ergs},
\end{eqnarray}
indicating that the total energy in electrons is much less than the energy
released through the spin-down of the pulsar. That both relic
electrons and wind electrons have roughly equal energy is consistent with the expectation given by \citet{1999A&A...346L..49A}.

\section{Cross calibration of IACTs \& Fermi}
\label{sec:crosscal}
The updated model of the SED of the Crab Nebula provides an opportunity for a
cross calibration between ground-based air shower experiments and the Fermi/LAT.
The method is demonstrated here with  the 
imaging air Cherenkov telescopes HEGRA, H.E.S.S., and MAGIC but is generally applicable to
any other experiment that measures the flux and spectrum
from the Crab Nebula in the high-energy regime.

In general, the energy calibration of IACTs is done indirectly with the help of detailed
simulations of air showers and the detector response. However, the remaining
systematic uncertainty on the absolute energy scale typically of 15\% leads to
substantial differences in the observed flux and position of cut-offs in the
energy spectra between different IACTs and also between Fermi/LAT and IACTs.
First attempts to cross-calibrate the IACTs among each other have already used the Crab Nebula
\citep{palaiseau2005...141,2008AIPC.1085..727Z}.  Since the known gamma-ray spectra lack sufficient sharp features, 
an absolute cross calibration with lines, etc., is not feasible. 
So far, efforts  to cross-calibrate Fermi/LAT with IACTs have focused on using the overlapping
energy range \citep{2005APh....23..572B}.
 Cross calibration
between Fermi/LAT and IACTs indirectly provides a means of benefiting
from the careful beam-line calibration of the Fermi/LAT \citep[see e.g.][]{2009ApJ...697.1071A}.

\begin{figure*}[tbh]
 \centering
\subfigure[The IC model (constant $B$-field model, solid black line) with
measurements from IACTs and Fermi/LAT. No energy scaling is applied.]{
 \label{fig:unscale}
 \includegraphics[width = 0.47\textwidth]{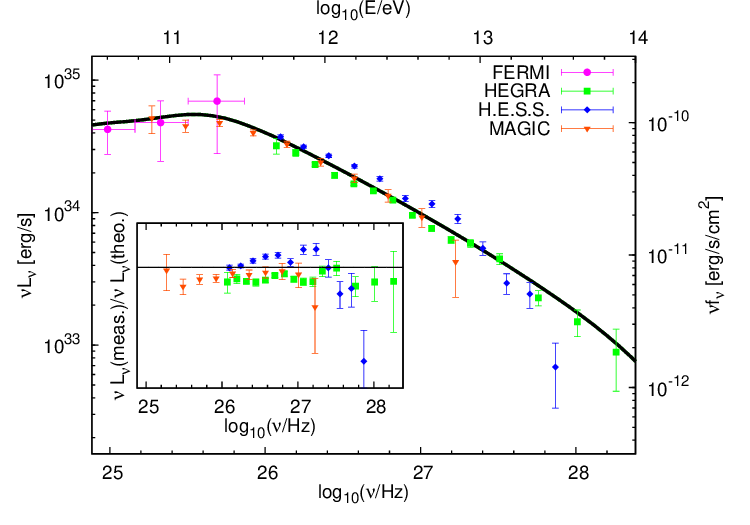}
}
 \hspace{10pt}
\centering
 \subfigure[The same curve as in \ref{fig:unscale} but with the
scaling factors of Eq. \ref{eqn:scale} and Table \ref{tbl:scale} applied to the data.]{
 \label{fig:scale}
 \includegraphics[width = 0.47\textwidth]{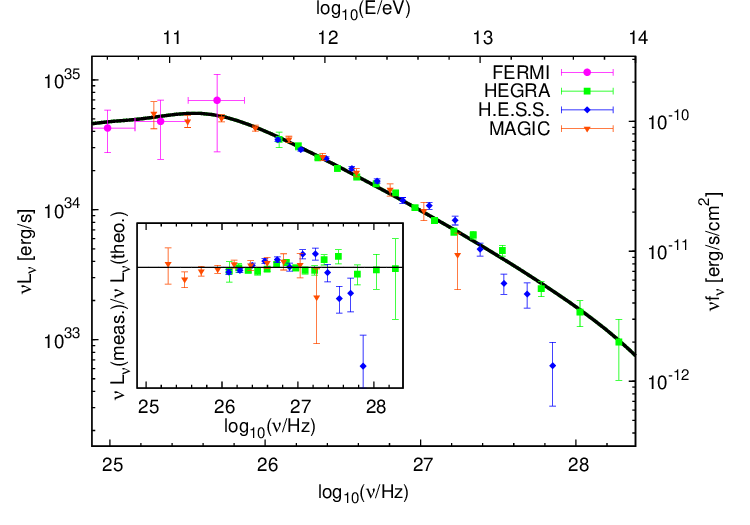}
 }
\caption[]{Comparison between the unscaled and scaled data of the IACTs.}
\end{figure*}

The cross calibration is accomplished in the following
way \citep{2009arXiv0912.3754M}: 
The average magnetic field of the model is adapted to
the Fermi observations.  
The statistical uncertainty on the $B$-field in Eq.~\ref{eqn:Bfit}
translates into statistical errors on this reference model (see Table \ref{tbl:scale}).
For each IACT an energy scaling factor
$s_\mathrm{IACT}$ is introduced to correct the measured energy $E_\mathrm{meas}$ to a common energy scale $E$ 
such that 
\begin{equation} E = E_\mathrm{meas} \cdot s_\mathrm{IACT}.
\label{eqn:scale} 
\end{equation} 
The scaling factor $s_\mathrm{IACT}$ for each instrument is determined via a
$\chi^2$-minimization in which the energy scale is changed according to
the formula above until the data points reproduce the model best. The resulting scaling
factors for the different instruments are listed in Table \ref{tbl:scale}
with the statistical errors and the reduced minimum $\chi^2$-values before and after the fit. 
To illustrate the result, Figs. \ref{fig:unscale} and
\ref{fig:scale} compare the
unscaled and the scaled data points  with the model. 
It is evident from these figures 
that the scaled data points fit the model better. This is also quantified by the $\chi^2$-values before and after the application of the scaling factors. All scaling factors lie within the aforementioned 15\% energy uncertainty of the IACTs.
The complete SED with the scaled data points and the model calculations of Section \ref{sec:SED} is shown in Appendix \ref{sec:fi-SED}.

The cross calibration eliminates the systematic uncertainty of the energy scale 
of the IACTs and adjusts them to a common one shared with the Fermi/LAT. 
However, the Fermi/LAT's absolute energy uncertainty remains, but it implies an improvement 
from $\pm 15 \%$ to $+5\%$ and $-10\%$. 

\begin{table*}
 \centering
 \begin{tabular}{ccccc} 
\textbf{Instrument}	& \textbf{Scaling factor $s_\mathrm{IACT}$}
& \textbf{Stat. error $\Delta s$}	& $\chi^2_\mathrm{before}\slash\mathrm{d.o.f.}$& $\chi^2_\mathrm{after}\slash\mathrm{d.o.f.}$\\
\hline
\hline
Fermi/LAT		& $1$	  	&$+0.05 ~-0.03$	&	--		&$0.49$		\\
HEGRA			& $1.042$ 	&$\pm 0.005$ 	&	$7.652$ 	&$1.046$		\\
H.E.S.S.		& $0.961$ 	&$\pm0.004 $ 	&	$11.84 $	&$6.476$		 \\
MAGIC			& $1.03$ 	&$\pm0.01  $ 	&	$1.671$		&$0.656$
 \end{tabular}
\caption{Energy scaling factors of the IACTs for the cross calibration.}
\label{tbl:scale}
\end{table*}

As a first application of the cross calibration, we derive upper limits
on the diffuse $\gamma$-ray background. Both Fermi \citep{2009PhRvL.102r1101A} and H.E.S.S. \citep{2008PhRvL.101z1104A,2009A&A...508..561A} have measured the cosmic ray $e^- + e^+$ spectrum.
Unlike Fermi/LAT, the telescopes from H.E.S.S. cannot accurately distinguish
between showers induced by electrons (or positrons) or photons, such that
up to $\approx 50$\% of the observed electromagnetic air showers
could be induced by photons. Therefore, H.E.S.S.
actually measures electrons and diffuse background photons. Taking the
difference of the two measurements we can derive an upper limit on the 
intensity of the $\gamma$-ray background. The scaling factors derived above are now
used to convert the H.E.S.S. data 
to the Fermi/LAT energy scale, which substantially reduces the systematic
uncertainty on the observed intensity, given that the electron spectrum follows a
soft power law with $\propto E^{-3}$. An important result of the cross calibration is that the peak in the spectrum
observed by ATIC \citep{2008Natur.456..362C} appears more unlikely after
applying the scaling factors.

\begin{figure}[tbh]
 \centering
 \includegraphics[angle=270, width = 0.48\textwidth]{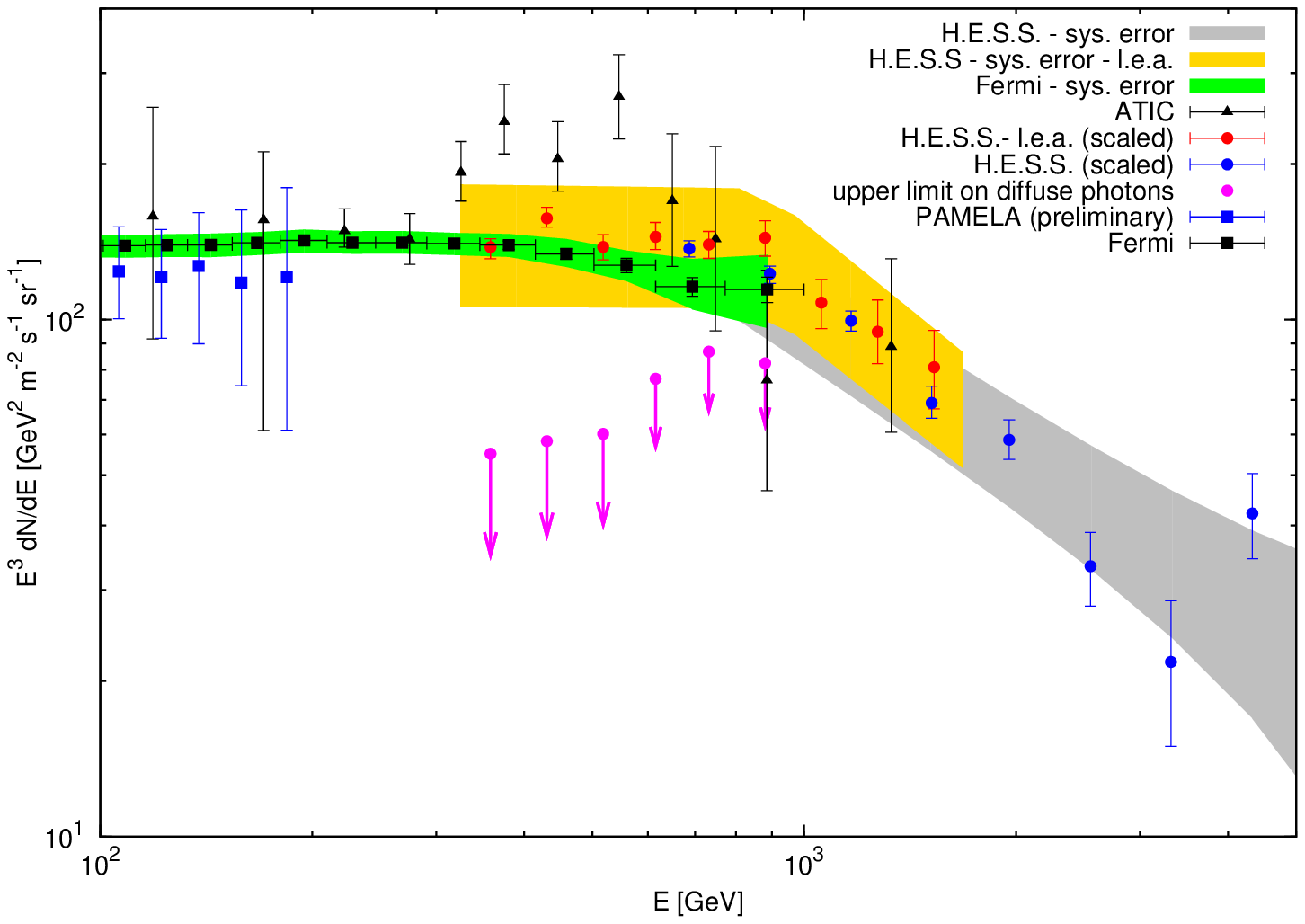}
 \caption{$e^- + e^+$ spectrum reported by H.E.S.S. and Fermi. The cross calibration factor
is applied to the H.E.S.S. energy scale, so the uncertainty on the global energy scale is reduced. The abbreviation l.e.a. stands for low-energy analysis as applied to the H.E.S.S. data in \cite{2009A&A...508..561A}. For comparison, the PAMELA data points are taken from \citet{Mocchiutti}.}
 \label{fig:atic}
\end{figure}

The upper limits are derived by taking the difference of the 
two measurements from the energy region covered by both instruments. This
corresponds to the first six H.E.S.S. points of the low energy analysis in
Fig. \ref{fig:atic}. The remaining systematic errors are taken into account for deriving the upper limits: the flux points of the H.E.S.S. measurements are shifted to
their maximum value allowed by the systematic uncertainties while the Fermi
points are shifted to the minimum value. This gives a 
conservative approximation for the upper limits. 
In general, the scaling factors can be used, e.g., to improve contraints on dark matter model parameters that rely on combined measurements
\citep[see e.g.][]{2010PhRvD..81j3521K}.

\section{Summary}
\label{sec:summary}

Updated models and data compilation for the SED of the Crab Nebula have been
presented.  The MHD flow model based upon KC84 and AA96 does not provide a
satisfactory description of the available data and requires refinements of the
underlying assumptions, e.g., relaxing 
spherical symmetry in axisymmetric numerical calculations as
carried out by Volpi et al. (2008). 
A straightforward modification is, among others, the
introduction of an anisotropic wind with variations of $\sigma$ when moving out
of the equatorial plane towards the polar regions of the outflow. The
simplified approach of a constant magnetic field pursued here has the benefit
of a smaller total number of parameters, even though the 
prescription of the injection spectrum in the MHD model is simpler 
(7 instead of 10 parameters) and physically more meaningful. 
Most important for the task of
cross-calibrating the instruments, the inverse Compton component predicted in
this model accurately describes all observational data above $1$~GeV for a
magnetic field of \bfield.  The comprehensive SED allows for an estimate of the
dust mass.  In contrast to \citet{2006AJ....132.1610T}, we obtain a value
about 40\% lower, namely $M_\mathrm{dust} = 4 \times 10^{-4} M_\odot$.  The
radio (relic) electrons provide additional (independent) constraints on the
average magnetic field in the entire radio nebula. Using the endpoint of the
electron spectrum the magnetic field is constrained to be smaller than $\approx
385~\mu$G, while the Fermi/LAT observations set a lower limit at approx.
$100~\mu$G.

The model describes the broad band data at high energies well enough to
derive energy scaling factors for cross-calibrating the Fermi/LAT instrument
with a variety of ground-based instruments. The cross-calibration eliminates
the systematic uncertainty of the different energy scales used by the IACTs and
ultimately reduces the global uncertainty to the Fermi/LAT calibration
uncertainties of $+5$\% and $-10$\%. An application of the cross calibration
were presented, and upper limits on the diffuse photon background between
200~GeV and 1~TeV derived by combining Fermi/LAT with H.E.S.S.
measurements. The excess measured by the ATIC collaboration seems unlikely
with the scaled H.E.S.S. observations. \\ The presented cross calibration can
be used universally when combining observational data from various instruments,
as well as in cases where the absolute energy scale is important.  Specifically
for soft spectra with cut-off features, such as in observations of objects at
cosmological distances suffering from absorption on the extragalactic
background light, the cross calibration can be useful for providing more stringent
upper limits.

\begin{appendix}

\section{Spatial distribution of seed photons and electrons in the nebula}
\label{sec:seed}
To calculate the seed photon fields required to
calculate the IC flux, it is neccessary to convert the observed photon flux to
the corresponding photon densities. This is done by following the approach
suggested by \citet{1998ApJ...503..744H}. Both the photon densities (apart from
the CMB contribution) and electron density are assumed to follow
Gaussian densities in the distance $r$ from the Nebula's center,
$\exp[-r^2\slash(2\sigma^2)]$ and $\exp[-r^2\slash(2\rho^2)]$, respectively.
The variances $\sigma^2$ and $\rho^2$ are energy dependent and estimated from
observations. For the Gaussian describing the photon density
\citet{1998ApJ...503..744H} find
 \begin{eqnarray}
 &{}&\arctan\left(\frac{\sigma}{d} \right) =\nonumber \\
 &{} &\left\{\begin{array}{ll}
                                          3.93\times10^{-4} & \mathrm{for}\, \epsilon < 0.02\unit{eV},\\
\\
					 \left(0.47 + 3.46\left[\frac{\epsilon}{0.02\unit{eV}}\right]^{-0.09}\right)\times10^{-4} & \mathrm{for}\, \epsilon \ge 0.02\unit{eV}.
                                         \end{array}\right.
\end{eqnarray}
Assuming that the photons are produced 
only via synchrotron radiation in a uniform magnetic field, 
the variance of the electron density follows directly from Eq. \ref{eqn:crit-freq} with the aforementioned averaging over $\theta$, 
\begin{eqnarray}
    &{}&\arctan\left(\frac{\rho}{d} \right) =  \nonumber\\
    &{}&\left\{\begin{array}{ll}
                                          3.93\times10^{-4} & \mathrm{for}\, \gamma < 6.65\times10^4\\ 
\\
					 \left(0.47 + 28.9~\gamma^{-0.17}\left[\frac{B}{10^{-4}\unit{G}}\right]^{-0.09}\right)\times10^{-4}& \mathrm{for}\, \gamma \ge 6.65\times10^4.\\
                                         \end{array}\right.\nonumber\\
\end{eqnarray}

Only those photons and electrons with overlapping distributions can interact, so that the total distribution is found by convolution to give a total variance of $\sigma^2 + \rho^2$. 
 For a photon production rate $S_\nu$, which is the sum of the contributions from synchrotron radiation, thermal dust emission, and optical line emission, we find
\begin{equation}
 n_\mathrm{Seed} = \frac{S_\nu}{4\pi c(\rho^2 + \sigma^2)} + n_\mathrm{CMB}.
\end{equation}
The photon production rates are calculated using the individual luminosities, $S_\nu = L_\nu\slash (h\nu)$, where
\begin{eqnarray}
L_\nu^\mathrm{sync}  & = &\int\limits_1^\infty  \Diff{\Nel}{\gamma} \mathcal{L}_\nu^\mathrm{Sy} \mathrm{d}\gamma,\\
  L_\nu^\mathrm{opt}  & = &\sum_i L_i^\mathrm{opt}\delta(\nu - \nu_i).
\end{eqnarray}
The fluxes from the line emissions are approximated by $\delta$-distributions, and the $L_i$ are given in the references, see Section \ref{sec:SED}. The luminosity for the dust emission is given by a gray body spectrum and the photon density of the CMB is calculated from a black body with a temperature of $T = 2.726$~K. The total seed photon field can then be used to calculate the inverse Compton emissions by means of Eqs. \ref{eqn:ic} and \ref{eqn:emission}.

\section{Single-particle emission functions}
\label{Appi}

The single-particle emission functions for particles of mass $m$ and energy $\gamma m c^2$ for synchrotron 
$\mathcal{L}_\nu^\mathrm{Sy}$ and inverse Compton emission $\mathcal{L}_\nu^\mathrm{IC}$
used in Eq.~\ref{eqn:emission} are taken from  \citet{1970RvMP...42..237B}: 
\begin{eqnarray}
 \mathcal{L}_\nu^\mathrm{Sy} &=& \frac{\sqrt 3 e^3 B \sin\theta}{m c^2}~\frac{\nu}{\nu_\mathrm{c}}\int\limits_{\nu\slash\nu_\mathrm{c}}^{~\infty} K_{5\slash 3} (x)~ \mathrm d x\, , \\ 
 \mathcal{L}_\nu^\mathrm{IC} &=& \frac 3 4 ~\frac{\sigma_\mathrm{T}c}{\gamma^2}h\nu\int\limits_{h\nu\slash(4\gamma^2)}^{~h\nu}~\mathrm d \epsilon ~\frac{n_\mathrm{seed}(\epsilon)}{\epsilon}f_\mathrm{IC}(\epsilon,\nu,\gamma).\label{eqn:ic} 
\end{eqnarray}
\textcolor{black}{
We denote the photon energy after scattering by $h \nu$ and before scattering by $\epsilon$, the Thomson cross section by $\sigma_\mathrm{T}$ and the electron charge by $e$. The critical frequency $\nu_\mathrm{c}$ is defined as 
}
\begin{equation}
\nu_\mathrm{c} = \frac{3 e}{4\pi m c}~B\sin\theta~\gamma^2,\label{eqn:crit-freq}
\end{equation}
and $K_{5\slash3}(x)$ stands for the modified Bessel function of fractional order $5\slash3$.
The electron pitch angle $\theta$ is averaged where the value $\langle\sin\theta\rangle=\sqrt{2/3}$ is adopted. Introducing the kinematic variable $q$,
\begin{eqnarray}
q = \frac{h \nu}{4\epsilon\gamma^2[1-h\nu\slash(\gamma m c^2)]},
\end{eqnarray}
the IC distribution function $f_\mathrm{IC}$ can be written as
\begin{eqnarray}
f_\mathrm{IC}(\epsilon,\nu,\gamma) & = & \nonumber\\
  2q\ln q & + & (1+2q)(1-q) + \frac 1 2~\frac{\left[{4\epsilon\gamma q}/{\left(mc^2\right)}\right]^2}{1+{4\epsilon\gamma q}/{\left(mc^2\right)}}(1-q).
\end{eqnarray}

\section{Correlation between the fit parameters}
\label{App:Cov}
The covariance matrix has been calculated at the position of the minimum in the $\chi^2$-function. The diagonal elements of the
covariance matrix have been already listed in the form of estimates of the error of the individual parameters, see Tables \ref{tbl:el_spec} and \ref{tbl:el_spec_mhd}. Here, we list
the correlation matrix defined as $\mathrm{cor}(i,j)=\sqrt{\mathrm{cov}(i,j)/(\sigma_i \sigma_j)}$ in Table \ref{tab:cov_const_B} for the constant $B$-field model and in Table \ref{tab:cov_mhd} for the MHD flow model. 
 \begin{table*}
\caption{The correlation coefficients between the fit-parameters for the constant B-field model. 
Since the matrix is symmetric, the lower trigonal part is not listed.
}
\label{tab:cov_const_B}
\begin{center}
\begin{tabular}{lrrrrrrrrrr}
    & $S_r$    & $\ln(N_0^r)$ & $\ln(\gamma^r_\mathrm{max})$ & $\ln(\gamma^w_\mathrm{min})$ & $\ln(\gamma^w_\mathrm{break})$ &
                                                                                                               $\ln(\gamma^w_\mathrm{max})$   &
													       $\beta$ &
                                                                                                               $S_w$  &
														$S_w+\Delta S$ &
														$\ln(N_0^w)$ \\
   \hline
   \hline
$S_r$                               & 1    &  $-1.00$ &  $-0.35$ & $-0.02$ & $-0.03$ & $-0.02$ &  $0.10$ & $0.05$ &  $0.02$ &  $0.03$   \\
$\ln(N_0^r)$                  &      &  1       &  $0.34$  & $0.02$  & $0.03$  & $0.02$  &  $-0.10$& $-0.05$ & $-0.01$ &  $-0.03$ \\
$\ln(\gamma^r_\mathrm{max})$  &      &          &  1       & $0.24$  & $-0.05$ & $-0.03$ &  $0.26$ & $0.08$  & $0.02$  &  $0.05$  \\
$\ln(\gamma^w_\mathrm{min})$   &      &          &          & 1       & $0.56$  & $ 0.13$ &  $-0.57$& $-0.80$ & $-0.25$ &  $-0.57$ \\
$\ln(\gamma^w_\mathrm{break})$&      &          &          &         &   1     & $0.39$  &  $-0.46$& $-0.82$ & $-0.70$ & $-1.00$  \\
$\ln(\gamma^w_\mathrm{max})$  &      &          &          &         &         &   1     &  $-0.11$& $-0.22$ & $-0.75$ & $-0.38$ \\
$\beta$                             &      &          &          &         &         &         &   1     & $0.71$  & $0.19$  & $0.47 $  \\
$S_w$                               &      &          &          &         &         &         &         &   1     & $0.42$  & $0.84$   \\
$S_w+\Delta S $                       &      &          &          &         &         &         &         &         &  1      & $0.69$   \\
$\ln(N_0^w)$                   &      &          &          &         &         &         &         &         &         &  1       \\
   \hline
\end{tabular}
\end{center}
\end{table*}
\begin{table*}
\caption{The correlation coefficients between the fit-parameters for the MHD flow model.
Since the matrix is symmetric, the lower trigonal part is not listed.
}
\label{tab:cov_mhd}
\begin{center}
 \begin{tabular}{lrrrrrrr}
     & $\ln(q_0)$ & $\ln(n_0)$ & $\ln(\gamma^r_\mathrm{max})$ & $\ln(\gamma^w_\mathrm{min})$ & $\ln(\gamma^w_\mathrm{max})$ & $S_w$ & $S_r$ \\
\hline \hline
$\ln(q_0)$                   &  1  & $-0.02$  &  $0.23$  &  $0.79$ &  $0.94$ & $-0.00$ &   $0.00$ \\
$\ln(n_0)$                   &     &  1       &  $-0.50$ & $-0.03$ &  $-0.02$&  $-0.01$  & $0.01$ \\
$\ln(\gamma^r_\mathrm{max})$ &     &          &        1 & $0.31$  &  $0.19$ &  $0.01$  & $-0.01$ \\
$\ln(\gamma^w_\mathrm{min})$ &     &          &          &      1  &  $0.65$ &  $0.01$  & $-0.01$ \\
$\ln(\gamma^w_\mathrm{max})$ &     &          &          &         &       1 &  $-0.11$  & $0.11$ \\
$S_w$                         &     &          &          &         &         &        1  & $-1.00$  \\
$S_r$                         &     &          &          &         &         &           & 1  \\
\hline
\end{tabular}
\end{center}
\end{table*}

\section{Parametrization of IC flux}
\label{sec:app-para}
For convenience, the energy flux $\nu f_\nu$ of the IC emission for $1$~GeV $< E < 1$~PeV is parametrized as a function of energy,
 namely a 5${}^\mathrm{th}$-order polynomial in double-logarithmic representation \citep[as done by][]{2004ApJ...614..897A}:
\begin{equation}
\log_{10}\left(\frac{\nu f_\nu}{\unit{erg}\unit{s}^{-1}\unit{cm}^{-2}}\right) = \sum_{i=0}^{5}p_i \log_{10}^i\left(\frac{E}{\unit{TeV}}\right).
\label{eqn:polynomial}
\end{equation}
The coefficients are given in Table \ref{tbl:polynomial}. The relative error of the parametrization for the parameters $p_3$ and $p_5$ are approximately 6\%, less than 1\% for $p_0$ and $p_2$ and about 1\% for $p_1$. The value of $p_4$ is set to zero since its relative error is otherwise around 150\%, and thus $p_4$ is not neccessary for a satisfactory fit. 

\begin{table}[tbh]
\caption{Parametrization of the IC flux. 
The coefficients correspond to Eq. \ref{eqn:polynomial}.}
\label{tbl:polynomial}
\centering
\begin{tabular}{ll}
Coefficient	&Value	\\
\hline
\hline
$p_0$~\ldots\ldots\ldots\ldots\ldots\ldots\ldots\ldots&	$-10.2708$ \\
$p_1$~\ldots\ldots\ldots\ldots\ldots\ldots\ldots\ldots&	$-0.53616$ \\
$p_2$~\ldots\ldots\ldots\ldots\ldots\ldots\ldots\ldots&	$-0.179475$ \\
$p_3$~\ldots\ldots\ldots\ldots\ldots\ldots\ldots\ldots&	$0.0473174$ \\
$p_4$~\ldots\ldots\ldots\ldots\ldots\ldots\ldots\ldots&	$0$ \\
$p_5$~\ldots\ldots\ldots\ldots\ldots\ldots\ldots\ldots&	$-0.00449161$
\end{tabular}
\end{table}

\section{Final SED}
\label{sec:fi-SED}

\begin{figure*}[htb]
\centering
\includegraphics[width= 1\linewidth]{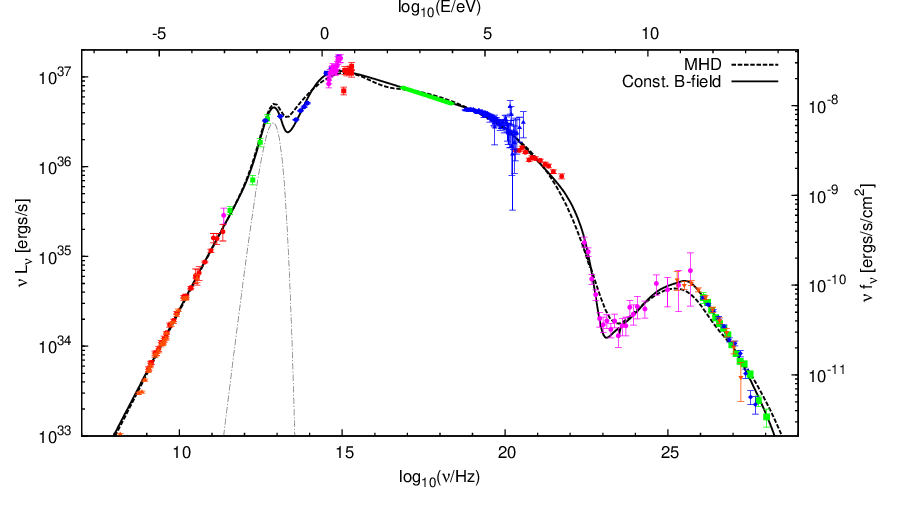}
\caption{The SED with the best-fitting model calculations.}
\label{fig:fi-SED}
\end{figure*}

Figure \ref{fig:fi-SED} summarizes the best fits for the constant B-field model and the MHD flow model (see Section \ref{sec:SED}), together with all data points of the references in Table \ref{tbl:data} and in \citet{2004ApJ...614..897A}. Likewise, the scaling factors for the IACTs introduced in Section \ref{sec:crosscal} are also applied.

\end{appendix}

\begin{acknowledgements}
This work was made possible with the support of the German federal ministry for
education and research (Bundesministerium f\"ur Bildung und Forschung) and the
collaborative research center (SFB) 676 ``Particle, Strings and the early
Universe'' at the University of Hamburg. We also like to thank the anonymous referee for useful comments. \end{acknowledgements}
\bibliographystyle{aa}
\bibliography{14108}

\end{document}